\documentclass[seceq]{ptptex}
\newcommand{\Slash}[1]{\ooalign{\hfil/\hfil\crcr$#1$}}
\title{%        
  Regularization of the Covariant Derivative on Curved Space
  
  by Finite Matrices
}

\author{%       %Use \scshape  for the family name
  Masanori \textsc{Hanada}\footnote{hana@gauge.scphys.kyoto-u.ac.jp}%
}

\inst{%         %Affiliation, neglected when [addenda] or [errata]
  Department of Physics, Kyoto University, 
  Kyoto 606-8502, Japan 
}

\abst{%         %this abstract is neglected when [addenda] or [errata]
  In a previous paper 
  [ M.~Hanada, H.~Kawai and Y.~Kimura, Prog. Theor. Phys. 114 (2005),
  1295 ] 
  it is shown that a covariant derivative on any 
  $n$-dimensional Riemannian manifold can be expressed in terms of  
  a set of $n$ matrices, and a new interpretation of IIB matrix model,  
  in which the diffeomorphism, the local Lorentz symmetry and their higher
  spin analogues are embedded in the unitary symmetry, is proposed. 
  In this article we investigate several coset manifolds in this
  formulation and show that on these backgrounds, 
  it is possible to carry out  
  calculations at the level of finite matrices  
  by using the properties of the Lie algebras. 
  We show how the local fields and the symmetries 
  are embedded as components of matrices 
  and how to extract the physical degrees of freedom 
  satisfying the constraint proposed in the previous paper. 
}

\begin{document}

\maketitle
%%%%%%%%%%%%%%%%%%%%%%%%%%%%%%%%%%%%%%%%%%%%%%%%%%%%%%% 
%%%%%%%%%%%%%%%%%%%%%%%%%%%%%%%%%%%%%%%%%%%%%%%%%%%%%%% 
\section{Introduction}
%%%%%%%%%%%%%%%%%%%%%%%%%%%%%%%%%%%%%%%%%%%%%%%%%%%%%%%
%%%%%%%%%%%%%%%%%%%%%%%%%%%%%%%%%%%%%%%%%%%%%%%%%%%%%%% 
Although it is conjectured that 
string theory provides 
the unification of fundamental interactions, 
its present formulation based on perturbation theory is not satisfactory. 
In order to examine whether it actually describes our four-dimensional world,
a non-perturbative and background independent formulation is needed. 
Matrix models represent a promising approach to studying the nonperturbative 
dynamics of string theory. 
For a critical string, 
they are basically obtained through dimensional reduction of 
the ten-dimensional $U(N)$ ${\cal N}=1$ supersymmetric Yang-Mills 
theory \cite{BFSS,IKKT,DVV}. 

IIB matrix model \cite{IKKT} 
is obtained through dimensional reduction to 
a point, and the action is given by 
\begin{eqnarray}
  S=-\frac{1}{g^2}Tr
  \left(
    \frac{1}{4}[A_{a},A_{b}][A^{a},A^{b}]
    +
    \frac{1}{2}\bar{\psi}\gamma^a[A_{a},\psi]
  \right), 
  \label{action_IIB}
\end{eqnarray}
where $\psi$ is a ten-dimensional Majorana-Weyl spinor, and 
$A_{a}$ and $\psi$ are $N\times N$ Hermitian matrices. 
The indices $a$ and $b$ are 
contracted by the flat metric. 
This action possesses an $SO(10)$ global Lorentz symmetry and $U(N)$ symmetry. 
One problem with this model is that 
it is unclear how curved spaces are described and 
how the fundamental principle of general relativity 
is realized in it. 

It may seem that the space of dynamical variables 
becomes very small after the dimensional reduction, 
but in fact this is not the case 
if $N$ is infinitely large \cite{EK}. 
Indeed, in Ref. \citen{HKK} a new interpretation,   
in which IIB matrix model contains 
gravity in a background independent manner, is given. 
The argument presented there is as follows. 
The matrix variables $A_{a}$ and $\psi_{\alpha}$ act on 
the Hilbert space $V={\mathbb C}^N$ as endomorphisms, i.e. 
linear maps from $V$ to itself. 
Because $V$ is infinite dimensional 
in the large-$N$ limit, we can give various different 
interpretations to it. 
If we assume $V$ is a space consisting of 
an $n$-component complex scalar field, i.e.,  
$V=\{\varphi^i:{\mathbb R}^{10}\to{\mathbb C}^n\}$,  
instead of $V={\mathbb C}^N$, 
an endomorphism $T$ is a bilocal field 
$K^{ij}(x,y)$, which can be formally regarded as 
the set of differential operators of arbitrary rank 
with $n\times n$ matrix coefficients \cite{hep-th/0204078}:
\begin{eqnarray}
  (T\varphi)^i(x)
  &=&
  \sum_{j=1}^n\int d^{10}y K^{ij}(x,y)\varphi^j(y)
  \nonumber\\
  &=&
  c_{(0)}^{ij}(x)\varphi^j(x)
  +
  c_{(1)}^{\mu,ij}(x)\partial_\mu\varphi^j(x)
  +
  c_{(2)}^{\mu\nu,ij}(x)\partial_\mu\partial_\nu\varphi^j(x)
  +
  \cdots. 
\end{eqnarray}
In particular, we can take the covariant derivative 
as a special value of $A_a$, 
\begin{eqnarray}
  A_a
  =i\left( \partial_a-ia_a(x) \right)
  \in End(V). 
\end{eqnarray}
In this sense, the ten-dimensional $U(n)$ ${\cal N}=1$ 
super Yang-Mills theory can be embedded 
in the matrix model, and local gauge symmetry is realized 
as a part of the $U(N)$ symmetry of the matrix model, 
\begin{eqnarray}
  \delta A_a=i[\lambda,A_a], 
\end{eqnarray}
where $\lambda$ is a matrix-valued function on ${\mathbb R}^{10}$, 
which is a $0$-th order differential operator in $End(V)$. 
With this interpretation of $V$, however, 
the model does not contain gravity. 
In order to embed gravity, we regard $A_a$ to be a covariant
derivative acting on a curved space. 
Then, the diffeomorphism and the local Lorentz symmetries become 
parts of the $U(N)$ symmetry of the matrix model. 
In this interpretation, 
any curved space corresponds to a certain matrix configuration,  
and the path integral includes 
the summation of all the curved spaces.    

The procedure for describing the covariant derivative with  
a set of matrices is the following \cite{HKK}.  
Let $M$ be an $n$-dimensional Riemannian manifold with a fixed 
spin structure and $M=\cup_i U_i$ be its open covering. 
On each patch $U_i$, the covariant derivative is expressed as  
\begin{eqnarray}
  \nabla_a^{[i]}
  =e_a{}^{m[i]}
  \left(
    \partial_m-\omega_m{}^{bc[i]}{\cal O}_{bc}
  \right), 
\end{eqnarray}
where $e_a{}^m$ and $\omega_m{}^{bc}$ are the vielbein 
and the spin connection, respectively. 
Here, ${\cal O}_{bc}$ is the Lorentz generator that acts on 
Lorentz indices. 
The index $[i]$ is the label of the patch. 
In the overlapping region $U_i\cap U_j$, 
$\nabla_a^{[i]}$ and $\nabla_a^{[j]}$ are related as 
\begin{eqnarray}
  \nabla_a^{[i]}
  =
  R_a{}^b(t_{ij}(x))\nabla_b^{[j]}, 
\end{eqnarray}
where $t_{ij}:U_i\cap U_j\to G=Spin(n)$ is the transition function  
and $R_a{}^b(t_{ij}(x))$ is the vector representation of $t_{ij}(x)$. 

Let us consider the principal $G$ bundle on $M$ associated 
with the spin structure, and denote it by $E_{prin}(M)$. 
It is constructed from the set of $U_i \times G$ 
by identifying 
$(x_{[i]},g_{[i]})$ with $(x_{[j]},g_{[j]})$:
\begin{eqnarray}
  x_{[i]}=x_{[j]}, 
  \qquad
  g_{[i]}=t_{ij}(x)g_{[j]}.
  \label{gluing condition}
\end{eqnarray}
We take $V=C^{\infty}(E_{prin}(M))$, which is the space of 
smooth functions on $E_{prin}(M)$. 
We assume that covariant derivatives act on the space $V$; 
that is, ${\cal O}_{ab}$ generates an infinitesimal left action, 
\begin{eqnarray}
  i\epsilon^{ab}
  \left(
    {\cal O}_{ab}f^{[i]}
  \right)(x,g)
  =
  f^{[i]}\left(
    x, \left(1+i\epsilon^{ab}M_{ab}\right)^{-1}g
  \right)
  -
  f^{[i]}\left(
    x,g
  \right),   
\end{eqnarray}
where $M_{ab}$ is the matrix of the fundamental representation. 
Then, 
we can construct endomorphisms from a covariant derivative 
as follows: 
\begin{eqnarray}
  \nabla_{(a)}^{[i]}=R_{(a)}{}^b(g_{[i]}^{-1})\nabla_b^{[i]}. 
\end{eqnarray}
Here, $R_{(a)}{}^b(g)$ is the vector representation of $G$.\footnote{
$R_{a}{}^b$ and $R_{(a)}{}^b$ are the same quantity. 
However, we formally distinguish them, because the $a$ and $(a)$ obey different 
transformation laws. 
Specifically, $a$ is transformed by the action of $G$, 
while $(a)$ is not. 
}
Each component of $\nabla_{(a)}$ is {\it globally} 
defined on $E_{prin}(M)$ 
(i.e. $\nabla_{(a)}^{[i]}=\nabla_{(a)}^{[j]}$ in an overlapping region) 
and is indeed an endomorphism on $V$. 
The index $(a)$ merely labels $n$ endomorphisms. 

This method is valid in any number of dimensions, and 
we can express the covariant derivative on any $n$-dimensional 
Riemannian manifold in terms of $n$ matrices.

Based on the procedure described above, the new interpretation of 
IIB matrix model \cite{HKK} mentioned above becomes feasible. 
If the matrices $A_a$ are sufficiently close to 
the covariant derivatives $i\nabla_{(a)}$  
on one of the manifolds $M$, 
it is natural to regard each $A_a$ as acting on
$C^\infty(E_{prin}(M))$    
and to expand $A_a$ about $i\nabla_{(a)}$:\footnote{
Strictly speaking, because $A_a$ is Hermitian, we should introduce  
the anticommutator $\{\ ,\ \}$ in 
Eq. (\ref{eq:expansion by local fields}):
\begin{eqnarray}
  A_a
  &=&
  i\nabla_{(a)}
  +
  a_{(a)}(x,g)
  +
  \frac{i}{2}\{a_{(a)}{}^{(b)}(x,g),\nabla_{(b)}\}
  +
  \frac{i}{2}\{a_{(a)}{}^{bc}(x,g),{\cal O}_{bc}\}
  +
  \cdots.   
\end{eqnarray}
} 
\begin{eqnarray}
  A_a
  &=&
  i\nabla_{(a)}
  +
  a_{(a)}(x,g)
  +
  ia_{(a)}{}^{(b)}(x,g)\nabla_{(b)}
  +
  ia_{(a)}{}^{bc}(x,g){\cal O}_{bc}
  \nonumber\\
  & &
  \quad
  +
  i^2a_{(a)}{}^{(b)(c)}(x,g)\nabla_{(b)}\nabla_{(c)}
  +
  i^2a_{(a)}{}^{(b),cd}(x,g)\nabla_{(b)}{\cal O}_{cd}
  +
  \cdots. 
  \label{eq:expansion by local fields}
\end{eqnarray}
In this expansion, local fields appear as coefficients.  
For example, $a_{(a)}{}^{(b)}(x,g)$ contains  
the fluctuations of the vielbein. 
Coefficients of higher-derivative terms 
correspond to fields of higher spin. 
In this sense, this part of the space of large-$N$ matrices 
describes the dynamics around the background spacetime $M$. 
The diffeomorphism and the local Lorentz 
symmetry are realized as parts of the $U(N)$ symmetry of 
the matrix model, which are generated by 
$\Lambda=\frac{i}{2}\{\lambda^{(a)},\nabla_{(a)}\}$ and 
$\Lambda=i\lambda^{ab}{\cal O}_{ab}$.\footnote{
Similar arguments have been given in the case of the spectral action 
\cite{CC,AST}. In this case, $V$ is taken to 
represent the spinor fields, and 
only bispinor operators, such as the Dirac operator, are considered. 
This is sufficient to include the symmetries of general relativity, 
although we cannot treat the action (\ref{action_IIB}) in this formalism. 
By comparing it with our formalism, 
it may possible to clarify how the physical degrees 
of freedom are embedded in matrices. 
} Note that {\it we do not fix} $M$ and 
{\it all Riemannian manifolds with all possible 
spin structures are included in the path integral}. 
In Ref. \citen{HKK2},  
this formalism is extended to the supermatrix model,  
so that the local supersymmetry is included in the superunitary
symmetry
\footnote{The same action has been studied  
in a different context\cite{OT}. }. 

Now let us consider the equation of motion \cite{HKK}.  
Variation with respect to $A_{a}$ gives the equation of motion 
\begin{eqnarray}
  \left[
    A^{a},[A_{a},A_{b}]
  \right]
  =0. 
  \label{eq:EOM with parenthesis_1}
\end{eqnarray}
Here we simply set $\psi=0$. 
If we impose the ansatz
\begin{eqnarray}
  A_{a}=i\nabla_{(a)}, 
  \label{ansatz:bosonic case}
\end{eqnarray}
Eq. (\ref{eq:EOM with parenthesis_1}) becomes
\begin{eqnarray}
  \left[
    \nabla^{(a)},[\nabla_{(a)},\nabla_{(b)}]
  \right]
  =0. 
  \label{eq:EOM with parenthesis_2}
\end{eqnarray}
Note that Eq. (\ref{eq:EOM with parenthesis_2}) is equivalent to
\begin{eqnarray}
  0
  &=&
  \left[
    \nabla^{a},[\nabla_{a},\nabla_{b}]
  \right]
  \nonumber\\
  &=&
  [\nabla^a,R_{ab}{}^{cd}{\cal O}_{cd}]
  \nonumber\\
  &=&
  \left(\nabla^a R_{ab}{}^{cd}\right){\cal O}_{cd}
  -
  R_b{}^c\nabla_c, 
  \label{eq:EOM without parenthesis}
\end{eqnarray}
where $R_{ab}{}^{cd}$ is the Riemann tensor and 
$R_b{}^c=R_{ab}{}^{ac}$ is the Ricci tensor. 
We have assumed that $\nabla_a$ is torsionless. 
Eq. (\ref{eq:EOM without parenthesis}) holds if and only if
\begin{eqnarray}
  \nabla^a R_{ab}{}^{cd}=0, 
  \quad
  R_{ab}=0. 
\end{eqnarray}
The first equation here follows from the second by the Bianchi identity 
$\nabla^{[a}R_{bc}{}^{de]}=0$. 
Therefore, the covariant derivative on a Ricci-flat spacetime 
is a classical solution.
Recently, it is also shown that the equation of motion 
of the massless higher spin gauge fields about flat space 
can be derived \cite{Saitou}. 

In general, the method presented in Ref. \citen{HKK} 
can be applied only in the large-$N$ limit, and 
we must introduce some regularization. 
In order to perform a Monte-Carlo simulation, 
or to calculate a divergent quantity like the free energy, 
an explicit regularization using finite-$N$ matrices is necessary.  
This is also useful for the purpose of extracting 
information concerning the topology from the matrices. 
The extensity of the eigenvalue distribution does not reveal  
the topology of the corresponding manifold; indeed, any manifold 
entails an infinitely large momentum.  
In order to determine the topology,  
we must investigate finer structures, e.g. commutation relations 
and the degeneracy of eigenvalues. 
On generic backgrounds, it is difficult to write 
the explicit forms of regularizations 
in terms of finite-$N$ matrices,  
but for certain manifolds with large symmetries, it is possible. 
In this paper, we give a few examples: spheres, real and complex 
projective spaces, tori and flat spaces. 

The organization of this paper is as follows. 
In \S \ref{sec:heat kernel} we discuss the heat kernel
regularization\footnote{This section is partially 
based on collaboration with H. Kawai and Y. Kimura.}. 
Although this method is not so useful in actual calculations, 
it can be applied to any background, 
and it makes clear the meaning of the regularizations presented in 
\S \ref{sec:regularization}. 
In \S \ref{sec:regularization} 
we introduce the explicit regularization procedures 
for spheres, real and complex projective spaces, tori and flat spaces. 
On these backgrounds, the corresponding $E_{prin}$ have the structures of 
Lie groups, and by using their algebraic properties, explicit 
calculations become tractable.   
We investigate the case of $S^2$ in detail and show how the local
fields and symmetries are embedded as components of matrices. 
Section \ref{sec:discussion} is devoted to conclusions. 
In Appendix \ref{appendix:heat kernel} we give 
formulae used in \S \ref{sec:heat kernel}. 
In Appendix \ref{sec:deformed action}, we discuss 
the classical solutions of modifications of IIB matrix model 
with mass or cubic terms. 
%%%%%%%%%%%%%%%%%%%%%%%%%%%%%%%%%%%%%%%%%%%%%%%%%%
%%%%%%%%%%%%%%%%%%%%%%%%%%%%%%%%%%%%%%%%%%%%%%%%%%
%%%%%%%%%%%%%%%%%%%%%%%%%%%%%%%%%%%%%%%%%%%%%%%%%%
\section{Heat kernel regularization}
\label{sec:heat kernel}
%%%%%%%%%%%%%%%%%%%%%%%%%%%%%%%%%%%%%%%%%%%%%%%%%%
%%%%%%%%%%%%%%%%%%%%%%%%%%%%%%%%%%%%%%%%%%%%%%%%%%
%%%%%%%%%%%%%%%%%%%%%%%%%%%%%%%%%%%%%%%%%%%%%%%%%%
In the previous section, we introduced the general procedure to 
express the covariant derivative in terms of a set of matrices. 
The Hilbert space $V=C^\infty(E_{prin}(M))$ 
is a space of smooth functions on $E_{prin}(M)$. 
In order to regularize the trace in a general covariant way, 
the heat kernel regularization \cite{Vassilevich} is useful. 

Let $\Delta_{prin}$ be the Laplacian on $E_{prin}(M)$, which is defined
by 
\begin{eqnarray}
  \Delta_{prin}
  =
  \sum_{(a)}\nabla_{(a)}{}^2
  +
  \kappa^2\sum_{a,b}{\cal O}_{ab}{}^2, 
\end{eqnarray}
where $\kappa$ is an arbitrary constant with dimensions of mass  
that specifies the mass scale in the direction of $Spin(n)$. 
As $\kappa$ becomes larger the damping factor on the higher
spin fields becomes more stringent.   
The heat kernel regularization is defined by
\begin{eqnarray}
\lefteqn{
  Tr X
  =
  \int dg \int e d^dx
  \langle x,g|X|x,g\rangle
}\nonumber\\
  &\longrightarrow&
  Tr_t X
  =
  \kappa^{d-D}
  \int dg \int e d^dx
  \langle x,g| e^{t\Delta_{prin}/2}
  X e^{t\Delta_{prin}/2}|x,g\rangle,  
\end{eqnarray}
where $t$ is a parameter with dimensions is $\mbox{(mass)}^{-2}$, 
$dg$ is the Haar measure of $Spin(n)$, 
$e=\det e_m{}^a$ is the determinant of the vielbein, 
and $D=\frac{d(d+1)}{2}$ is the dimension of $E_{prin}(M)$.  
Eventually we take the limit $t\to 0$.   
Using the heat kernel regularization, we can cut off  
the higher frequency modes of $C^\infty(E_{prin}(M))$. 
In terms of the fields on $M$,  
the modes with large momentum or large spin\footnote{
Note that the momentum along the direction of $Spin(n)$ correspond 
to the spin. } are suppressed.  
Expanding the matrix variables in local fields as 
Eq. (\ref{eq:expansion by local fields}) and using 
the heat kernel regularization, 
we can express physical quantities in terms of the local fields. 
As an example, let us evaluate the bosonic part of the action. 
If we ignore the local fields other than the graviton, 
then using a formula given in Appendix \ref{appendix:heat kernel},  
it becomes 
\begin{eqnarray}
  S_{bosonic}
  &=&
  -\frac{1}{4g^2}
  Tr_t
  \left(
    [i\nabla_{(a)},i\nabla_{(b)}]^2
  \right)
  \nonumber\\
  &=&
  -\frac{1}{4g^2}
  Tr_t
  \left(
    R_{(a)(b)}{}^{cd}{\cal O}_{cd}
  \right)^2
  \nonumber\\
  &=&
  \frac{1}{4g^2}
  \frac{\kappa^{d-D-2}t^{-D/2-1}}{(4\pi)^{D/2}}
  \int dg\int ed^dx
  R_{abcd}R^{abcd}
  +
  O(t^{-D/2-2}). 
\end{eqnarray}
If we include the higher spin fields, their kinetic terms and 
nonlinear interactions involving derivatives appear. 
For example, consider the case in which  
the spin $3$ field $a_a{}^{mn}(x)$ is excited. 
For simplicity, let us assume that the background spacetime is flat. 
Then we have 
\begin{eqnarray}
  S_{bosonic}
  &=&
  -\frac{1}{4g^2}
  Tr_t
  \left(
    \left[i\partial_{(a)}
      +
      \frac{i^2}{2}
      \left\{
        a_{(a)}{}^{(c)(d)},\partial_{(c)}\partial_{(d)}
      \right\},
      i\partial_{(b)}
      +
      \frac{i^2}{2}
      \left\{
        a_{(b)}{}^{(e)(f)},\partial_{(e)}\partial_{(f)}
      \right\}
    \right]^2
  \right)
  \nonumber\\
  &=&  
  \frac{1}{4g^2}
  \frac{\kappa^{d-D}t^{-D/2-3}}{4(4\pi)^{D/2}}
  \int dg\int ed^dx
  \Bigl(
  a_a{}^{cd}(\partial_c a_a{}^{ef})
  a_a{}^{c^\prime d^\prime}(\partial_c a_b{}^{e^\prime f^\prime})
  \nonumber\\
  & & 
  \hspace{6cm}
  -
  a_a{}^{cd}(\partial_c a_a{}^{ef})
  a_b{}^{c^\prime d^\prime}(\partial_c a_a{}^{e^\prime f^\prime})     
  \Bigl)
  \nonumber\\
  & &
  \qquad\times
  \left(
    \delta_{dd^\prime}\delta_{ee^\prime}\delta_{ff^\prime}
    +
    \mbox{14 permutations}
  \right)
  +
  O(t^{-D/2-2}). 
\end{eqnarray}
The kinetic term of $a_a{}^{bc}(x)$ appears at $O(t^{-D/2-2})$. 
In general, the action becomes far more complicated,  
as an infinite number of higher spin fields appear 
and interact with each other. 
Couplings with the curvature of the background spacetime also exist. 

Some remarks are in order here. 
First, although the heat kernel regularization preserves 
the general covariance, it breaks the higher spin gauge symmetries. 
Because the higher spin symmetries are necessary for the
consistency of the higher spin gauge theory, 
it is desirable to find a regularization which does not break them.  
Secondly, because the heat kernel regularization 
depends explicitly on the background spacetime, 
it is not suitable for a nonperturbative study. 
For such a purpose, a regularization by finite-$N$ matrices, in which 
we can perform calculations directly in terms of matrices, is desirable. 
Thirdly, although the heat kernel regularization does not 
give the Einstein-Hilbert action, the Ricci-flat spacetimes are 
classical solutions, as we saw in the introduction. 
This is possible because in the derivation of the equation 
of motion of the matrix model, 
Eq. (\ref{eq:EOM with parenthesis_1}) 
[or Eq. (\ref{eq:EOM with parenthesis_2})],   
we varied not only the graviton but also 
all the fields with any spin. 
This can easily be done in terms of matrices, 
but once we rewrite the matrix model in terms of the local fields, 
such a calculation becomes hopelessly complicated; 
calculations become difficult already at the classical level.  
%%%%%%%%%%%%%%%%%%%%%%%%%%%%%%%%%%%%%%%%%%%%%%%%%%
%%%%%%%%%%%%%%%%%%%%%%%%%%%%%%%%%%%%%%%%%%%%%%%%%%
%%%%%%%%%%%%%%%%%%%%%%%%%%%%%%%%%%%%%%%%%%%%%%%%%%
\section{Regularization by finite-$N$ matrices}
\label{sec:regularization}
%%%%%%%%%%%%%%%%%%%%%%%%%%%%%%%%%%%%%%%%%%%%%%%%%%
%%%%%%%%%%%%%%%%%%%%%%%%%%%%%%%%%%%%%%%%%%%%%%%%%%
%%%%%%%%%%%%%%%%%%%%%%%%%%%%%%%%%%%%%%%%%%%%%%%%%%
In the previous section we discussed the heat kernel regularization. 
Although it can be applied 
to any background spacetime and has an apparent physical interpretation, 
the actual calculation seems to be almost impossible,  
because in the regularized action, an infinite number of higher spin
fields couple in a complicated manner. 
Furthermore, because it depends explicitly on the background spacetime, 
it is not suitable when we consider nonperturbative dynamics, 
e.g. the dynamical generation of the spacetime.  
These facts motivate us to introduce a regularization 
by finite-$N$ matrices.  

In general, writing an explicit form of such a regularization is difficult,  
but there are classes of manifolds which have large symmetries, 
so that the corresponding $E_{prin}$ have Lie group structures and allow 
explicit finite-$N$ regularizations 
that can be obtained using representation theory. 
In this section, we present a few examples: 
$S^n,\ {\mathbb R}P^n,\ {\mathbb C}P^n,\ T^n$ and ${\mathbb R}^n$. 
We explain the case of $S^2$ in detail and show how the local fields 
and the generators of the diffeomorphism, the local Lorentz
transformation, and the higher spin gauge transformations 
are embedded in the matrices. 
Similar arguments can also be applied to other backgrounds. 
%%%%%%%%%%%%%%%%%%%%%%%%%%%%%%%%%%%%%%%%%%%%%%%%%%
%%%%%%%%%%%%%%%%%%%%%%%%%%%%%%%%%%%%%%%%%%%%%%%%%%
%%%%%%%%%%%%%%%%%%%%%%%%%%%%%%%%%%%%%%%%%%%%%%%%%%
\subsection{The covariant derivative on $S^n$}\label{subsec:sphere}
%%%%%%%%%%%%%%%%%%%%%%%%%%%%%%%%%%%%%%%%%%%%%%%%%%
%%%%%%%%%%%%%%%%%%%%%%%%%%%%%%%%%%%%%%%%%%%%%%%%%%
%%%%%%%%%%%%%%%%%%%%%%%%%%%%%%%%%%%%%%%%%%%%%%%%%%
Let us consider the 
$n$-sphere $S^n$ with an isotropic and homogeneous metric. 
The commutation relations of the covariant derivatives 
$\nabla_{(a)}\ (a=1,\cdots,n)$ are given by 
\begin{eqnarray}
  [\nabla_{(a)},\nabla_{(b)}]
  =
  \frac{R}{n(n-1)}
  \left\{
    \delta_{a}{}^c
    \delta_{b}{}^d
    -
    \delta_{a}{}^d
    \delta_{b}{}^c
  \right\}{\cal O}_{cd}^{\langle n\rangle}
  =
  \frac{2R}{n(n-1)}{\cal O}_{ab}^{\langle n\rangle}, 
\end{eqnarray}
where $R$ is the scalar curvature. 
Here ${\cal O}_{ab}^{\langle n\rangle}$ 
is the generator of $Spin(n)$, 
which satisfies the commutation relation 
\begin{eqnarray}
  [{\cal O}_{ab}^{\langle n\rangle},
  {\cal O}_{cd}^{\langle n\rangle}]
  =
  \frac{1}{2}
  \left\{
    \delta_{ac}{\cal O}_{bd}^{\langle n\rangle}
    -
    \delta_{bc}{\cal O}_{ad}^{\langle n\rangle}
    -
    \delta_{ad}{\cal O}_{bc}^{\langle n\rangle}
    +
    \delta_{bd}{\cal O}_{ac}^{\langle n\rangle}
  \right\}. 
\end{eqnarray}
It acts on $\nabla_{(a)}$ as 
\begin{eqnarray}
  [{\cal O}_{bc}^{\langle n\rangle},
  \nabla_{(a)}]
  =
  \frac{1}{2}
  \left\{
    \delta_{ab}\nabla_{(c)}
    -
    \delta_{ac}\nabla_{(b)}
  \right\}. 
\end{eqnarray}
Therefore, the algebra generated by $\nabla$ 
and ${\cal O}^{\langle n\rangle}$ is equivalent to $spin(n+1)$ 
under the identification 
\begin{eqnarray}
  \nabla_{(a)}
  \longleftrightarrow
  2\sqrt{\frac{R}{n(n-1)}}{\cal O}^{\langle n+1\rangle}_{a,n+1}, 
  \qquad
  {\cal O}^{\langle n\rangle}_{ab}
  \longleftrightarrow
  {\cal O}^{\langle n+1\rangle}_{ab}, 
  \label{correspondence between nabla and generators of spin(n+1)}
\end{eqnarray}
where ${\cal O}^{\langle n+1\rangle}_{ab}$ is the generator of $Spin(n+1)$. 
Therefore, $E_{prin}$ corresponding to $S^n$ is $Spin(n+1)$: 
\begin{eqnarray}
  E_{prin}\left(S^n\right)=Spin(n+1). 
\end{eqnarray}
This result is in some sense trivial, because it is well known that 
\begin{eqnarray}
  S^n=Spin(n+1)/Spin(n). 
\end{eqnarray}
Here, $Spin(n)$ is interpreted as the local Lorentz degrees of freedom. 

In order to express the covariant derivative in terms of a set of matrices, 
we must take the Hilbert space to consist of the smooth functions on 
$E_{prin}\left(S^n\right)=Spin(n+1)$, 
which is the regular representation, by definition. 
It is decomposed as 
\begin{eqnarray}
  C^\infty\left(Spin(n+1)\right)
  =
  \oplus_{r:{\rm irr.repr.}}
  \underbrace{
    \left(
      V_r\oplus\cdots\oplus V_r
    \right)}_{\mbox{dim} V_r\ \mbox{times}},  
  \label{regular representation:irreducible decomposition}
\end{eqnarray}
where in the r.h.s., $r$ runs over all the irreducible representations of 
$Spin(n+1)$. 
As we explain for the case of $S^2$ in detail below, 
by discarding the larger representations, 
we obtain a sensible cutoff. 
This is an analogue of the heat kernel regularization, 
because the Casimir operator of $spin(n+1)$ is 
simply the Laplacian on $E_{prin}(S^n)=Spin(n+1)$. 
%%%%%%%%%%%%%%%%%%%%%%%%%%%%%%%%%%%%%%%%%%%%%%%%%%
%%%%%%%%%%%%%%%%%%%%%%%%%%%%%%%%%%%%%%%%%%%%%%%%%% 
%%%%%%%%%%%%%%%%%%%%%%%%%%%%%%%%%%%%%%%%%%%%%%%%%% 
\subsubsection{The covariant derivative on $S^2$}
%%%%%%%%%%%%%%%%%%%%%%%%%%%%%%%%%%%%%%%%%%%%%%%%%%
%%%%%%%%%%%%%%%%%%%%%%%%%%%%%%%%%%%%%%%%%%%%%%%%%% 
%%%%%%%%%%%%%%%%%%%%%%%%%%%%%%%%%%%%%%%%%%%%%%%%%%
\noindent
\textbf{The covariant derivative and a truncation of 
  the Hilbert space}\\ 
%%%%%%%%%%%%%%%%%%%%%%%%%%%%%%%%%%%%%%%%%%%%%%%%%%
%%%%%%%%%%%%%%%%%%%%%%%%%%%%%%%%%%%%%%%%%%%%%%%%%% 
%%%%%%%%%%%%%%%%%%%%%%%%%%%%%%%%%%%%%%%%%%%%%%%%%%
\hspace{0.51cm}
As a concrete example, let us consider the case of $n=2$ in detail. 
In this case, we can identify $\nabla$ and ${\cal O}$ with 
the generators $J_i$ of $Spin(3)=SU(2)$ as 
\begin{eqnarray}
  i\nabla_{(1)}
  \longleftrightarrow
  \sqrt{\frac{R}{2}}J_1, 
  \qquad
  i\nabla_{(2)}
  \longleftrightarrow
  \sqrt{\frac{R}{2}}J_2, 
  \qquad
  i{\cal O}^{\langle 2\rangle}_{12}
  \longleftrightarrow
  \frac{1}{2}J_3, 
  \label{nabla and O:S2}
\end{eqnarray}
where generators $J_i$ satisfy the commutation relations 
\begin{eqnarray}
  [J_i,J_j]=i\epsilon_{ijk}J_k. 
\end{eqnarray}

Let us label the states in the Hilbert space 
by the ``angular momentum" $(l,m)$ as 
\begin{eqnarray}
  {\mathbb J}^2|l,m;i\rangle
  =
  l(l+1)|l,m;i\rangle, 
  \qquad
  J_3|l,m;i\rangle
  =
  m|l,m;i\rangle. 
\end{eqnarray}
Here, $i=-l,-l+1,\cdots,l$ is the label 
for the different spin $l$ representations 
in the regular representation\footnote{
  We use $i=-l,\cdots,l$ instead of $1,\cdots,2l+1$,  
  because it is related to the ``angular momentum'',  
  as we see below.
}. 
In our interpretation, the eigenvalue $m$ of $J_3$ is the spin of the state. 
This state has momentum $\sqrt{\frac{R}{2}\left(l(l+1)-m^2\right)}$, because 
\begin{eqnarray}
  -\nabla^2|l,m;i\rangle
  =
  \frac{R}{2}\left(
    {\mathbb J}^2-J_3^2
  \right)|l,m;i\rangle
  =
  \frac{R}{2}\left(l(l+1)-m^2\right)|l,m;i\rangle. 
\end{eqnarray}
For each spin and momentum, there are $2l+1$ states labeled by $i$ . 
To understand the reason for this degeneracy, let us consider the case of 
a scalar field, $m=0$. 
A scalar field on a sphere can be expanded in spherical harmonics. 
In terms of spherical harmonics, 
the state with momentum $l(l+1)$ is 
the state with the angular momentum $l$, 
which has $2l+1$ degrees of freedom. 

To see the correspondence discribed above more directly, 
it is instructive to recall the proof of the irreducible decomposition 
(\ref{regular representation:irreducible decomposition}). 
Let $G$ be a compact Lie group. 
[~In the present case, we have $G=Spin(3)$.~]  
We use 
the orthonormality of the representation matrices, expressed as 
\begin{eqnarray}
  \frac{1}{Vol(G)}\int dgR_{ij}^{\langle r\rangle}(g)^{\ast}
  R_{kl}^{\langle r^{\prime}\rangle}(g)=\frac{1}{d_{r}}
  \delta^{\langle r\rangle \langle r^{\prime}\rangle}
  \delta_{ik}\delta_{jl}, 
\end{eqnarray}
where $dg$ is the Haar measure, 
$R_{ij}^{\langle r\rangle}(g)$ is the representation matrix 
for the irreducible representation $r$, and $d_r={\rm dim}V_r$ is 
the dimension of the representation $r$. 
By the Peter-Weyl theorem, the representation matrices 
$R_{ij}^{\langle r\rangle}(g)$ form a complete set of 
smooth functions from 
$G$ to ${\mathbb C}$. 
Therefore, any smooth function $f(g)$ can be expanded as 
\begin{equation}
  f(g)=\sum_{r}c_{ij}^{\langle r\rangle}
  \sqrt{d_{r}}R_{ij}^{\langle r\rangle}(g), 
  \label{expansionoff(g)}
\end{equation}
where the sum is taken over all irreducible representations. 
We now see how the quantities $c_{ij}^{\langle r\rangle}$ 
transform under the action of $G$. 
The right-hand side transforms as 
\begin{eqnarray}
  \hat{h}:c_{ij}^{\langle r\rangle}R_{ij}^{\langle r\rangle}(g)
  &\rightarrow&
  c_{ij}^{\langle r\rangle}R_{ij}^{\langle r\rangle}(h^{-1}g) 
  =
  c_{ij}^{\langle r\rangle}R_{ik}^{\langle r\rangle}(h^{-1})
  R_{kj}^{\langle r\rangle}(g), 
\end{eqnarray} 
which shows that $c_{ij}^{\langle r\rangle}$ transforms as 
\begin{eqnarray}
  c_{ij}^{\langle r\rangle}
  \rightarrow
  ^{T}\!\!R_{ik}^{\langle r\rangle}(h^{-1})c_{kj}^{\langle r\rangle}. 
\end{eqnarray}
The index $i$ of $c_{ij}$ transforms 
as the dual representation 
of $r$, while the index $j$ is invariant. It follows that 
$c_{ij}^{\langle r\rangle}$ represents $d_{r}$ copies of 
the dual representation of $r$. 
Thus, we have found that the regular representation is decomposed as 
in Eq. (\ref{regular representation:irreducible decomposition}). 
In the present case, the state $|l,m;i\rangle$ corresponds to 
$\sqrt{2l+1}R^{\langle l\rangle}_{mi}(g)$. 
Now consider the global right action 
\begin{eqnarray}
  \hat{r}(h):f(g)\mapsto f(gh), 
  \label{right action of SU(2)}
\end{eqnarray}
which commutes with the left action and acts on
$E_{prin}(S^2)=Spin(3)$ as a rotation. 
[ Note that such a transformation
exists only when $E_{prin}$ is a Lie group.]   
Under this transformation, the index $i$ of 
$|l,m;i\rangle$ is rotated as follows:
\begin{eqnarray}
  \hat{r}(h):
  |l,m;i\rangle
  \mapsto
  R_{ji}^{\langle l\rangle}(h)|l,m;j\rangle. 
\end{eqnarray}
Therefore, the states $|l,0;i\rangle\ (i=-l,-l+1,\cdots,l)$ correspond to 
the spherical harmonics with angular momentum $(l,i)$, 
while the states  
$|l,m;i\rangle$ are the corresponding modes of spin $m$ fields. 

Next let us consider how to express the covariant derivative 
$\nabla_{(a)}$ in terms of a set of finite-$N$ matrices. 
As explained above, the smaller representations correspond to 
the states with the smaller momenta and the lower spins. 
Therefore, it is natural to cut off the larger
representations\footnote{
This can be regarded as a finite-$N$ realization of the heat kernel 
regularization. }. 
If we discard the representations whose spins are larger than $L$, 
we have the representation 
\begin{eqnarray}
  V_L
  &=&
  \underset{l\le L,\ l:{\rm half\ integer}}{\oplus}
  \underbrace{
    \left(
      V_l\oplus\cdots\oplus V_l
    \right)}_{2l+1\ \mbox{times}}
  \nonumber\\
  &=&
  Span\left\{
    |l,m;i\rangle\ \biggl| 
    l=0,\frac{1}{2},\cdots,L;\ m,i=-l,\cdots,l
  \right\}, 
\end{eqnarray}
whose dimension is 
\begin{eqnarray}
  N
  =\sum_{0\le l\le L,\ l\in {\mathbb Z}/2}(2l+1)^2
  =\frac{1}{24}(4L+2)(4L+3)(4L+4). 
\end{eqnarray}
Because the momentum of the state $|l,m;i\rangle$ is 
$\sqrt{\frac{R}{2}\left(l(l+1)-m^2\right)}$, 
the momentum cutoff becomes more stringent as the spin $m$ increases. 
The covariant derivatives can be expressed by the rotation generators
acting on this space. In the following,   
we denote the projection to $V_L$ by $\hat{\pi}_L$, 
but we omit it for simplicity, unless it is confusing.  
%%%%%%%%%%%%%%%%%%%%%%%%%%%%%%%%%%%%%%%%%%%%%%%%%%
%%%%%%%%%%%%%%%%%%%%%%%%%%%%%%%%%%%%%%%%%%%%%%%%%% 
%%%%%%%%%%%%%%%%%%%%%%%%%%%%%%%%%%%%%%%%%%%%%%%%%%
\\ \\
\textbf{Diffeomorphism, local Lorentz symmetry and local fields}\\ 
%%%%%%%%%%%%%%%%%%%%%%%%%%%%%%%%%%%%%%%%%%%%%%%%%%
%%%%%%%%%%%%%%%%%%%%%%%%%%%%%%%%%%%%%%%%%%%%%%%%%% 
%%%%%%%%%%%%%%%%%%%%%%%%%%%%%%%%%%%%%%%%%%%%%%%%%%
\hspace{0.51cm}
Because we are using a rather simple basis, 
diffeomorphisms and local Lorentz transformations 
can be written explicitly in terms of components.  

First, let us consider the operator $\hat{f}$, 
which acts on the Hilbert space $C^\infty(Spin(3))$  
as a multiplication by $f(x,g)$:\footnote{
  Although in the introduction we did not distinguish $f$ and $\hat{f}$, 
  here we distinguish them 
  so that $\hat{f}$ is not misinterpreted as  
  an element of the Hilbert space. 
}
\begin{eqnarray}
  \hat{f}:h(x,g)\mapsto f(x,g)\cdot h(x,g).  
\end{eqnarray}
Here, $x$ and $g$ denote the coordinates of $M$ and $Spin(2)=U(1)$,
respectively. 
Such an operator can be written as a linear combination of 
$\hat{R}_{mn}^{\langle l\rangle}$ in the form 
\begin{eqnarray}
  \hat{f}=\sum_{l,m,n}f^{\langle l\rangle}_{mn}
  \cdot\sqrt{2l+1}\hat{R}_{mn}^{\langle l\rangle}, 
\end{eqnarray}
where 
\begin{eqnarray}
  f(g)=\sum_{l,m,n}f^{\langle l\rangle}_{mn}
  \cdot\sqrt{2l+1}R_{mn}^{\langle l\rangle}(x,g). 
\end{eqnarray}
Because a product of representation matrices satisfies the relation 
\begin{eqnarray}
\lefteqn{
  \sqrt{2l+1}R_{mn}^{\langle l\rangle}(x,g)
  \cdot
  \sqrt{2k+1}R_{ij}^{\langle k\rangle}(x,g)
}\nonumber\\
  &=&
  \sum_{L=|k-l|}^{k+l}
  C(L;l,m,n;k,i,j)\cdot
  \sqrt{2L+1}R_{m+i,n+j}^{\langle L\rangle}(x,g), 
\end{eqnarray}
where $C(L;l,m,n;k,i,j)$ is a product of Clebsch-Gordan coefficients,
given by  
\begin{eqnarray}
\lefteqn{
  C(L;l,m,n;k,i,j)
}\nonumber\\
  &=&
  \sqrt{\frac{(2l+1)(2k+1)}{2L+1}}
  \left(
    \langle L,l,k,m+i|l,m\rangle|k,i\rangle
  \right)^\ast
  \langle L,l,k,n+j|l,n\rangle|k,j\rangle, 
\end{eqnarray}
$\hat{R}_{mn}^{\langle l\rangle}$ can be written as 
\begin{eqnarray}
  \sqrt{2l+1}\hat{R}_{mn}^{\langle l\rangle}
  =
  \sum_{L,k,i,j}
  C(L;l,m,n;k,i,j)
  |L,m+i;n+j\rangle
  \langle k,i;j|. 
\end{eqnarray}
Because we are using the basis in which the covariant derivative 
and the Lorentz generator are block-diagonal, 
a multiplication of functions is basically off-diagonal. 
Note that $\hat{R}_{mn}^{\langle l\rangle}$ raises the spin of 
the state by $m$, and hence it can be regarded as an operator of spin
$m$.\footnote{
  In Ref. \citen{HKK}, it is shown that the field $f^{\langle
    r\rangle}_{(a)}(x,g)$ 
  of representation $r$ transforms under 
  the right action 
  $\hat{r}(h):f(x,g)\mapsto f(x,gh)$
  of $Spin(n)$ as 
  \begin{eqnarray}
    \hat{r}(h^{-1})\hat{f}^{\langle
      r\rangle}_{(a)}\hat{r}(h)
    =
    R_{(a)}^{\langle r\rangle(b)}(h)
    \hat{f}^{\langle r\rangle}_{(b)}. 
  \end{eqnarray}
  In the present case, because $Spin(2)=U(1)$ is abelian, 
  the left and right actions are the same, 
  and it is generated by the adjoint action of $J_3$. 
  Therefore, $\hat{R}^{\langle l\rangle}_{mn}$ has spin $m$ 
  also in this definition. 
  Note that the right action of $Spin(2)$ considered here is 
  different from that of $Spin(3)$, defined by 
  Eq. (\ref{right action of SU(2)}). 

} 
 
Next, let us consider the local Lorentz transformation. 
Because a  parameter $\lambda(x)=\lambda^{01}(x)$ 
of the local Lorentz transformation is 
a scalar field on $S^2$, it can be expanded in  
$R_{0n}^{\langle l\rangle}$:
\begin{eqnarray}
  \lambda(x)
  =
  \sum_{l,n}
  \lambda^{\langle l\rangle}_{n}
  \cdot\sqrt{2l+1}
  R_{0n}^{\langle l\rangle}(x), 
  \qquad
  \hat{\lambda}
  =
  \sum_{l,n}
  \lambda^{\langle l\rangle}_{n}
  \cdot\sqrt{2l+1}
  \hat{R}_{0n}^{\langle l\rangle}. 
\end{eqnarray}
Using this, we can write the generator of the local Lorentz
transformation as 
\begin{eqnarray}
  \Lambda=\hat{\lambda}J_3. 
\end{eqnarray}

In the case of the diffeomorphism, 
a parameter $\lambda^{(a)}(x,g)$ is a vector on $S^2$, 
and therefore it can be expanded in $R_{\pm 1,n}^{\langle l\rangle}$:
\begin{eqnarray}
  \lambda^{(\pm 1)}(x,g)
  &=&
  \sum_{l,n}
  \lambda^{(\pm 1)\langle l\rangle}_{n}
  \cdot\sqrt{2l+1}
  R_{\pm 1,n}^{\langle l\rangle}(x,g), 
  \nonumber\\
  \hat{\lambda}^{(\pm 1)}
  &=&
  \sum_{l,n}
  \lambda^{(\pm 1)\langle l\rangle}_{n}
  \cdot\sqrt{2l+1}
  \hat{R}_{\pm 1,n}^{\langle l\rangle}. 
\end{eqnarray}
By using $J_{\pm}=J_1\pm iJ_2$, the generator of 
the diffeomorphism becomes 
\begin{eqnarray}
  \Lambda
  =
  \frac{1}{2}\sqrt{\frac{R}{2}}
  \left\{
    \hat{\lambda}^{(1)},J_{-}
  \right\}
  +
  \frac{1}{2}\sqrt{\frac{R}{2}}
  \left\{
    \hat{\lambda}^{(-1)},J_{+}
  \right\}.  
\end{eqnarray}
Similarly, we can express all the higher spin gauge transformations
explicitly in terms of matrices. 

When we regularize the Hilbert space $C^\infty(Spin(3))$ 
using $V_L$, we must restrict $\Lambda$ to act only on $V_L$:
\begin{eqnarray}
  \Lambda
  \longrightarrow
  \Lambda_L
  \equiv
  \hat{\pi}_L
  \Lambda
  \hat{\pi}_L. 
\end{eqnarray}
Note that this projection does not commute with the multiplication, 
\begin{eqnarray}
  \left(\Lambda\Lambda^\prime\right)_L
  \neq
  \Lambda_L\Lambda_L^\prime, 
\end{eqnarray}
but a pathology exists only near the momentum cutoff. 

The expansion in local fields given in 
Eq. (\ref{eq:expansion by local fields}) is expressed 
in terms of matrices, just as in the case of $\Lambda$. 
%%%%%%%%%%%%%%%%%%%%%%%%%%%%%%%%%%%%%%%%%%%%%%%%%%
%%%%%%%%%%%%%%%%%%%%%%%%%%%%%%%%%%%%%%%%%%%%%%%%%% 
%%%%%%%%%%%%%%%%%%%%%%%%%%%%%%%%%%%%%%%%%%%%%%%%%%
\\ \\ \textbf{Perturbative instability and its possible 
  elimination}\\ 
%%%%%%%%%%%%%%%%%%%%%%%%%%%%%%%%%%%%%%%%%%%%%%%%%%
%%%%%%%%%%%%%%%%%%%%%%%%%%%%%%%%%%%%%%%%%%%%%%%%%% 
%%%%%%%%%%%%%%%%%%%%%%%%%%%%%%%%%%%%%%%%%%%%%%%%%%
\hspace{0.51cm}
Once the regularization is specified, we can calculate various quantities. 
Unfortunately, however, it is not possible to carry out 
a perturbative expansion about this background, 
because of the presence of the tachyonic modes, 
which come from the higher spin fields. 
This can be seen as follows. 
First, we decompose the matrix variables 
into the background and the fluctuations, as
\begin{eqnarray}
  A_a=P_a+a_a, 
  \qquad
  \psi_\alpha=0+\chi_\alpha, 
\end{eqnarray}
where $P_a=i\nabla_{(a)}=\sqrt{\frac{R}{2}}J_a$ for $a=1,2$  
and $P_a=0$ otherwise, 
and add the gauge fixing and ghost terms 
$S_{GF+FP}=-Tr\left(\frac{1}{2}[P_a,a_a]^2+[P_a,b][A_a,c]\right)$. 
Then, the kinetic terms become 
\begin{eqnarray}
  S_{kin}
  =
  \frac{1}{g^2}Tr\ 
  \left\{
    \frac{1}{2}a_a\left({\cal P}^2\delta_{ab}-2i{\cal F}_{ab}\right)a_b
    -
    \frac{1}{2}\bar{\chi}\Slash{{\cal P}}\chi
    +
    b{\cal P}^2c
  \right\}, 
  \label{kinetic term}
\end{eqnarray}
where ${\cal P}_a=[P_a,\ \cdot\ ]$ and 
${\cal F}_{ab}=i[{\cal P}_a,{\cal P}_b]$. 
In the present case, ${\cal P}^2=\frac{R}{2}\left({\cal J}_1^2+{\cal J}_2^2\right)$ and 
${\cal F}_{12}=-\frac{R}{2}{\cal J}_3$, 
where ${\cal J}_i=[J_i,\ \cdot\ ]$, 
commute and hence are simultaneously diagonalizable. 
The adjoint representation is labeled by the matrix indices 
$(l,m;i)$ and $(k,n;j)$, and each of them decomposes into 
the direct sum of $(2k+1)(2l+1)$ copies of 
spin $|l-k|,\cdots,l+k$ representations. 
These bases are related by the Clebsch-Gordan coefficients as 
\begin{eqnarray}
  X_{(l,m;i)(k,n;j)}
  \longleftrightarrow
  \tilde{X}_{(s,p;i,j)}
  =
  \sum_{l,m;k,n}
  X_{(l,m;i)(k,n;j)}
  \times
  \langle l,k,s,p
  |l,m\rangle
  |k,n\rangle^\prime,  
\end{eqnarray}
where $|k,n\rangle^\prime$ is 
the dual representation of $|k,n\rangle$. 
Here, $\tilde{X}_{(s,p;i,j)}$ is a spin $(s,p)$ state of ${\cal J}$, 
which corresponds to a field of spin $p$ \footnote{
  Strictly speaking, $p$ represents the $g$-dependence of 
  $a_{(a)}{}^{bcd\cdots}(x,g)$, which is obtained from 
  $a_{(a)}{}^{(b)(c)(d)\cdots}(x,g)$ in Eq. (\ref{eq:expansion by local fields}) 
  by changing the indices with parentheses to those 
  without: 
  \begin{eqnarray}
    a_{(a)}{}^{bcd\cdots}(x,g)
    =
    R_{b}{}^{(b^\prime)}(g)
    R_{c}{}^{(c^\prime)}(g)R_{d}{}^{(d^\prime)}(g)
    \cdots
    a_{(a^\prime)}{}^{(b^\prime)(c^\prime)(d^\prime)\cdots}(x,g). 
  \end{eqnarray}
}
and momentum 
$\sqrt{\frac{R}{2}\left(s(s+1)-p^2\right)}$.

A straightforward but tedious calculation shows 
that adjoint representation we are considering 
can be decomposed as 
\begin{eqnarray}
  \oplus_{s\le 2L}
  \underbrace{
    \left(
      V_s\oplus\cdots\oplus V_s
    \right)}_{k_s\ \mbox{times}}, 
\end{eqnarray}
where 
\begin{eqnarray}
  k_s
  =
  \left\{
    \begin{array}{cc}
      \frac{1}{6}
      \left(
        s-2L-1
      \right)
      \left(
        2s^3+6s^2+4Ls^2-20sL-5s-16L^2s-8L^2-14L-6
      \right), & (s=\mbox{integer})\\
      \frac{1}{48}(2s+1)(2s+8L+7)(2s-4L-1)(2s-4L-3). & (s=\mbox{odd}/2) 
    \end{array}
  \right.
  \nonumber\\
\end{eqnarray}
For simplicity, we have assumed here that $L$ is an integer. 
Note that the dimension is 
$\sum_{s=0}^{2L}(2s+1)k_s=
\left(
  \frac{1}{24}(4L+2)(4L+3)(4L+4)
\right)^2=N^2$, 
as expected. 
In this basis, we have  
\begin{eqnarray}
  \left({\cal P}^2\delta_{ab}-2i{\cal F}_{ab}\right)
  \sim
  \frac{R}{2}
  \left(
    \begin{array}{cccc}
      s(s+1)-p^2 & -2ip & & \\
      2ip & s(s+1)-p^2 & & \\
      & & s(s+1)-p^2 & \\
      & & & \ddots
    \end{array}
  \right)_{ab}, 
  \nonumber\\
\end{eqnarray}
but this is not positive-definite, because 
tachyons arise in the case 
\begin{eqnarray}
  \left(s(s+1)-p^2\right)^2-4p^2<0.  
\end{eqnarray}  
Even if we introduce an infrared cutoff (that is, we discard 
small $s$),  
this inequality holds if $p$ is sufficiently large. 

Some remarks are in order here. 
By adding a tachyonic mass term, 
\begin{eqnarray}
  S_{mass}=-\frac{R}{2ng^2}
  Tr\left(A_aA^a\right), 
\end{eqnarray}
we can make an $n$-sphere with scalar curvature $R$ a classical solution
\footnote{In general, spaces satisfying the Einstein equation 
  with cosmological constant 
  $\frac{n-2}{2n}R$ become classical solutions. For details, see
  Appendix \ref{sec:deformed action}. }. 
Therefore, if such a term is generated dynamically, 
then the spacetime would compactify spontaneously. 
However, there remain some subtleties. 
First, the addition of such a term does not remove the tachyonic modes. 
Secondly, a fuzzy sphere and fuzzy torus also become classical solutions. 
The tree-level free energies on these backgrounds are 
of order $-\frac{R}{g^2}N^3$, which are far smaller than 
those of ordinary spheres, $\sim -\frac{R}{g^2}N^{5/3}$. 
Therefore, the existence of tachyonic modes is not so surprising.   
If an ordinary sphere became stable in the large-$N$
limit, then the dynamical correction would be very important:
The higher spin fields should acquire mass dynamically and 
decouple from the low energy spectrum. 
It would interesting to study whether or not this is the case 
nonperturbatively, 
for example by employing a Monte Carlo simulation, as in 
Ref. \citen{ABNN}. 
It would also be interesting to find other modifications of the action 
that make curved spaces classical solutions. 
Another way to remove the tachyons is to require $A_{a}$ 
to transform as a vector or scalar\footnote{
  Since $a=3,4,\cdots,10$ should be regarded as a label of scalar 
  on $S^2$, it is natural to require the second constraint 
  in Eq. (\ref{transformation under right action}). 
}  under the global right action of $Spin(2)$: 
\begin{eqnarray}
  \hat{r}(h^{-1})A_a\hat{r}(h)
  &=&
  R_a{}^b(h)A_b\ (a=1,2), 
  \nonumber\\
   \hat{r}(h^{-1})A_a\hat{r}(h)
  &=&
  0\ (a=3,4,\cdots,10).  
  \label{transformation under right action}
\end{eqnarray}
Under this assumption, 
$a_{a}{}^{bcd\cdots}(x,g)$, which is obtained from 
$a_{(a)}{}^{(b)(c)(d)\cdots}(x,g)$ in Eq. (\ref{eq:expansion by local fields}) 
by 
\begin{eqnarray}
  a_{a}{}^{bcd\cdots}(x,g)
  =
  R_{a}{}^{(a^\prime)}(g)R_{b}{}^{(b^\prime)}(g)
  R_{c}{}^{(c^\prime)}(g)R_{d}{}^{(d^\prime)}(g)
  \cdots
  a_{(a^\prime)}{}^{(b^\prime)(c^\prime)(d^\prime)\cdots}(x,g), 
\end{eqnarray}
does not depend on $g$ \cite{HKK}. 
Because $Spin(2)=U(1)$ is
abelian, the left and right actions are the same, and it is generated 
by the adjoint action of $J_3$. Therefore, imposing the above
condition, the modes with $|p|>1$ are removed. 
Note that the modes removed here are those which do not allow 
simple interpretations in terms of low energy field theory. 
Although in the present case the constraint 
(\ref{transformation under right action}) can easily be written 
in terms of matrices as
\begin{eqnarray}
  [J_3,A_{\pm 1}]=\pm A_{\pm 1} 
  \ (A_{\pm}=A_1\pm iA_2), 
  \qquad
  [J_3,A_a]=0\ (a=3,4,\cdots,10), 
\label{constraint S2}
\end{eqnarray}
on gereric backgrounds it is difficult to write 
this condition explicitly. 
Note that such a condition depends explicitly on the background. 
In any case, if we remove tachyonic modes by modifing the action 
or imposing the condition (\ref{transformation under right action}), 
we can perform a perturbative calculation around it. 
As we saw above, the calculation in this case is almost parallel to that 
on a fuzzy sphere.  As an example, we calculate the one-loop free 
energy under the constraint (\ref{transformation under right action}). 
%%%%%%%%%%%%%%%%%%%%%%%%%%%%%%%%%%%%%%%%%%%%%%%%%%
%%%%%%%%%%%%%%%%%%%%%%%%%%%%%%%%%%%%%%%%%%%%%%%%%% 
%%%%%%%%%%%%%%%%%%%%%%%%%%%%%%%%%%%%%%%%%%%%%%%%%%
\\ \\ \textbf{One-loop free energy under the constraint 
  (\ref{constraint S2})}\\ 
%%%%%%%%%%%%%%%%%%%%%%%%%%%%%%%%%%%%%%%%%%%%%%%%%%
%%%%%%%%%%%%%%%%%%%%%%%%%%%%%%%%%%%%%%%%%%%%%%%%%% 
%%%%%%%%%%%%%%%%%%%%%%%%%%%%%%%%%%%%%%%%%%%%%%%%%%
\hspace{0.51cm}
We impose on $\psi,b$ and $c$ 
conditions similar to the constraint (\ref{constraint S2}) \cite{HKK}, 
\begin{eqnarray}
  [J_3,\psi]=\pm\frac{1}{2}\gamma^{12}\psi, 
  \qquad
  [J_3,b]=[J_3,c]=0. 
  \label{constraint S2:fermion and ghosts}
\end{eqnarray}
Then, from Eq. (\ref{kinetic term}), the one-loop free energy becomes 
\begin{eqnarray}
  F_{one\mbox{-}loop}
  &=&
  \frac{1}{2}Tr_{b}
  \log\left({\cal P}^2\delta_{ab}-2i{\cal F}_{ab}\right)
  -
  \frac{1}{2}Tr_{f}
  \log{\Slash{{\cal P}}}
  -
  Tr_{gh}{\cal P}^2, 
  \label{eq:free energy}
\end{eqnarray}
where the subscripts on $Tr$ indicate that we take the trace 
only over the states satisfying Eqs. 
(\ref{constraint S2}) and (\ref{constraint S2:fermion and ghosts}). 
Each term in Eq. (\ref{eq:free energy}) is evaluated as follows:
\begin{eqnarray}
  \lefteqn{
    \frac{1}{2}Tr_{b}
    \log\left({\cal P}^2\delta_{ab}-2i{\cal F}_{ab}\right)
  }\nonumber\\
  &=&
  5\sum_{s:{\rm integer}}k_s
  \log \frac{R}{2}s(s+1)
  +
  \frac{1}{2}\sum_{s:{\rm integer}}k_s
  \log\left\{
    \frac{
      \left(s(s+1)-1\right)^2-4
    }{s^2(s+1)^2}
  \right\}, 
  \nonumber\\ \\
  -
  \frac{1}{2}Tr_{f}
  \log{\Slash{{\cal P}}}
  &=&
  -\frac{1}{4}Tr_f
  \log\left({\cal P}^2+\frac{i}{2}\Slash{{\cal F}}\right)
  \nonumber\\
  &=&
  -2
  \sum_{s={\rm odd}/2}k_s
  \log \frac{R^2}{4}\left\{
    \left(
      s(s+1)-\frac{1}{4}
    \right)^2
    -\frac{1}{4}
  \right\}, 
  \\
  -Tr_{gh}{\cal P}^2
  &=&
  -\sum_{s:{\rm integer}}k_s
  \log \frac{R}{2}s(s+1). 
\end{eqnarray}
Therefore, we find 
\begin{eqnarray}
  F_{one\mbox{-}loop}
  &=&
  4\sum_{s:{\rm integer}}k_s\log s(s+1)
  -2
  \sum_{s={\rm odd}/2}k_s
  \log\left\{
    \left(
      s(s+1)-\frac{1}{4}
    \right)^2
    -\frac{1}{4}
  \right\}
  \nonumber\\
  & &
  \quad
  +
  \frac{1}{2}\sum_{s:{\rm integer}}k_s
  \log\left\{
    \frac{
      \left(s(s+1)-1\right)^2-4
    }{s^2(s+1)^2}
  \right\}. 
\end{eqnarray}
In order to remove tachyonic modes completely, 
we must introduce an infrared cutoff; that is, 
we should discard the modes satisfying $s\le 1$ by hand. 
Then, numerically we find that it behaves as 
\begin{eqnarray}
  F_{one\mbox{-}loop}
  \sim
  -L^{4}
  \sim
  -N^{4/3},  
\end{eqnarray}
and is negligible compared with the tree-level free energy.  
%%%%%%%%%%%%%%%%%%%%%%%%%%%%%%%%%%%%%%%%%%%%%%%%%%
%%%%%%%%%%%%%%%%%%%%%%%%%%%%%%%%%%%%%%%%%%%%%%%%%% 
%%%%%%%%%%%%%%%%%%%%%%%%%%%%%%%%%%%%%%%%%%%%%%%%%%
\\ \\ \textbf{Difference from the fuzzy sphere}\\ 
%%%%%%%%%%%%%%%%%%%%%%%%%%%%%%%%%%%%%%%%%%%%%%%%%%
%%%%%%%%%%%%%%%%%%%%%%%%%%%%%%%%%%%%%%%%%%%%%%%%%% 
%%%%%%%%%%%%%%%%%%%%%%%%%%%%%%%%%%%%%%%%%%%%%%%%%%
\hspace{0.51cm}
Here we stress again that although $\nabla$ is expressed 
in terms of the generators of 
$SU(2)$, it is completely different from the fuzzy sphere. 
We can express the ordinary two-sphere using two matrices, 
while in the case of the fuzzy two-sphere, three matrices are needed. 
Because the covariant derivative on the ordinary sphere 
is only a part of $su(2)$, 
it is not a classical solution of the
matrix model with the cubic term \cite{CS}, 
although the fuzzy sphere is a classical solution. 
(~For details, see Appendix \ref{sec:deformed action}.~)
The radii are also different; 
the radii of the commutative and fuzzy spheres are 
$\sim\frac{1}{\sqrt{R}}$ and 
$\frac{N}{\sqrt{R}}$, respectively. 
%%%%%%%%%%%%%%%%%%%%%%%%%%%%%%%%%%%%%%%%%%%%%%%%%%
%%%%%%%%%%%%%%%%%%%%%%%%%%%%%%%%%%%%%%%%%%%%%%%%%% 
%%%%%%%%%%%%%%%%%%%%%%%%%%%%%%%%%%%%%%%%%%%%%%%%%% 
\subsection{The covariant derivative on ${\mathbb R}P^n$}
%%%%%%%%%%%%%%%%%%%%%%%%%%%%%%%%%%%%%%%%%%%%%%%%%%
%%%%%%%%%%%%%%%%%%%%%%%%%%%%%%%%%%%%%%%%%%%%%%%%%% 
%%%%%%%%%%%%%%%%%%%%%%%%%%%%%%%%%%%%%%%%%%%%%%%%%%
${\mathbb R}P^n$ is obtained from $S^n$ by 
identifying antipodal points. 
The corresponding principal bundle $E_{prin}({\mathbb R}P^n)$ 
is also obtained through this identification, and so we have 
\begin{eqnarray}
E_{prin}({\mathbb R}P^n)
=
Spin(n+1)/\pm
=
SO(n+1). 
\end{eqnarray}
This can be seen also from the relation 
\begin{eqnarray}
SO(n+1)/Spin(n)
=
{\mathbb R}P^n. 
\end{eqnarray}
As in the case of $Spin(n+1)$, 
$C^\infty(SO(n+1))$ can be expanded in the irreducible representations 
of $SO(n+1)$, and the restriction to smaller representations 
provides a natural cutoff, which is an analogue of the heat kernel
regularization. 
%%%%%%%%%%%%%%%%%%%%%%%%%%%%%%%%%%%%%%%%%%%%%%%%%%
%%%%%%%%%%%%%%%%%%%%%%%%%%%%%%%%%%%%%%%%%%%%%%%%%% 
%%%%%%%%%%%%%%%%%%%%%%%%%%%%%%%%%%%%%%%%%%%%%%%%%% 
\subsubsection{The covariant derivative on ${\mathbb R}P^2$}
%%%%%%%%%%%%%%%%%%%%%%%%%%%%%%%%%%%%%%%%%%%%%%%%%%
%%%%%%%%%%%%%%%%%%%%%%%%%%%%%%%%%%%%%%%%%%%%%%%%%% 
%%%%%%%%%%%%%%%%%%%%%%%%%%%%%%%%%%%%%%%%%%%%%%%%%%
Functions on $SO(3)$ can be expanded in the representation matrices
with integer spins:
\begin{eqnarray}
  C^\infty\left(SO(3)\right)
  &=&
  \underset{l:{\rm integer}}{\oplus}
  \underbrace{
    \left(
      V_l\oplus\cdots\oplus V_l
    \right)}_{2l+1\ \mbox{times}}
  \nonumber\\
  &=&
  Span\left\{
    |l,m;i\rangle\ \biggl| 
    l=0,1,2,\cdots;\ m,i=-l,-l+1,\cdots,l
  \right\}
  \nonumber\\
  &=&
  Span\left\{
    \left.
      R_{mi}{}^{\langle l\rangle}
    \right|l:\mbox{integer};\ m,i=-l,-l+1,\cdots,l
  \right\}. 
\end{eqnarray}
By restricting the value of $l$, we can introduce 
an explicit finite-$N$ regularization, 
\begin{eqnarray}
  V_L=
  \underset{l\le L,\ l:{\rm integer}}{\oplus}
  \underbrace{
    \left(
      V_l\oplus\cdots\oplus V_l
    \right)}_{2l+1\ \mbox{times}}
  =
  Span\left\{
    |l,m;i\rangle\ \biggl| 
    l=0,\frac{1}{2},\cdots,L;\ m,i=-l,\cdots,l
  \right\}, 
  \nonumber\\
\end{eqnarray}
where
\begin{eqnarray}
  N=\frac{1}{6}(2L+1)(2L+2)(2L+3). 
\end{eqnarray}
The covariant derivative is expressed in $SO(3)$ generators 
acting on this space, just as in Eq. (\ref{nabla and O:S2}). 
The only difference from the case of $S^2$ 
is that in the present case, only 
the integer spins appear\footnote{
  $R_{ij}{}^{\langle l\rangle}$ is symmetric (resp., antisymmetric) 
  under the identification of the antipodal points of $Spin(3)\simeq
  S^3$ if $l\in{\mathbb Z}_{\ge 0}$ 
  (resp., $l\in{\mathbb Z}_{\ge 0}+\frac{1}{2}$). 
  Functions on ${\mathbb R}P^3=SO(3)$ are the functions on $S^3=Spin(3)$ 
  that are symmetric under this identification. 
}. 
Therefore, the local fields and various symmetries 
are embedded in the same way, 
and if we expand the matrices about this background, 
tachyonic modes appear, due to the existence of higher spin 
fields, just as in the case of $S^2$. 
%%%%%%%%%%%%%%%%%%%%%%%%%%%%%%%%%%%%%%%%%%%%%%%%%%
%%%%%%%%%%%%%%%%%%%%%%%%%%%%%%%%%%%%%%%%%%%%%%%%%% 
%%%%%%%%%%%%%%%%%%%%%%%%%%%%%%%%%%%%%%%%%%%%%%%%%% 
\subsection{The covariant derivative on ${\mathbb C}P^n$}
%%%%%%%%%%%%%%%%%%%%%%%%%%%%%%%%%%%%%%%%%%%%%%%%%%
%%%%%%%%%%%%%%%%%%%%%%%%%%%%%%%%%%%%%%%%%%%%%%%%%% 
%%%%%%%%%%%%%%%%%%%%%%%%%%%%%%%%%%%%%%%%%%%%%%%%%%
In the case of ${\mathbb C}P^n$, the covariant derivative 
and the Lorentz generators form a subalgebra of $spin(2n+2)$, 
and a procedure similar to that explained above can be applied.  
Although $E_{prin}\left({\mathbb C}P^n\right)$ is not 
so simple, if we are interested only in holomorphic quantities 
then the covariant derivatives on ${\mathbb C}P^n$ takes 
simple forms. 
Using the indices $J$ and $\bar{J}$, defined by 
\begin{eqnarray}
  v^J=v^{2j-1}+\sqrt{-1}v^{2j}, 
  \qquad
  v^{\bar{J}}=v^{2j-1}-\sqrt{-1}v^{2j},   
\end{eqnarray}
under the identification 
\begin{eqnarray}
  \nabla_{(I)}
  \leftrightarrow
  2\sqrt{\frac{R}{n(n+1)}}{\cal O}_{I,\overline{n+1}}^{\langle
    2n+2\rangle}, 
  \quad
  \nabla_{(\bar{I})}
  \leftrightarrow
  2\sqrt{\frac{R}{n(n+1)}}{\cal O}_{n+1,\bar{I}}^{\langle
    2n+2\rangle},   
  \quad
  {\cal O}_{I\bar{J}}^{\langle
    2n\rangle}
  \leftrightarrow
   {\cal O}_{I\bar{J}}^{\langle
    2n+2\rangle},  
    \nonumber\\
\end{eqnarray}
and
\begin{eqnarray}
  \sum_{I=1}^{n+1}
  {\cal O}_{I\bar{I}}^{\langle 2n+2\rangle}
  =
  0, 
\end{eqnarray}
the algebra generated by $\nabla$ becomes $su(n+1)$. 
Therefore, taking $C^\infty\left(SU(n+1)\right)$ as a Hilbert space, 
the covariant derivative can be expressed 
in terms of the rotation generators acting on it. 
$C^\infty\left(SU(n+1)\right)$ can be decomposed 
into the irreducible representations of $SU(n+1)$, and 
regularization by finite-$N$ matrices is realized by cutting off 
the larger representations.  
%%%%%%%%%%%%%%%%%%%%%%%%%%%%%%%%%%%%%%%%%%%%%%%%%% 
%%%%%%%%%%%%%%%%%%%%%%%%%%%%%%%%%%%%%%%%%%%%%%%%%% 
%%%%%%%%%%%%%%%%%%%%%%%%%%%%%%%%%%%%%%%%%%%%%%%%%% 
\subsection{The covariant derivative on ${\mathbb R}^n$ and $T^n$}\label{subsec:flat}
%%%%%%%%%%%%%%%%%%%%%%%%%%%%%%%%%%%%%%%%%%%%%%%%%%
%%%%%%%%%%%%%%%%%%%%%%%%%%%%%%%%%%%%%%%%%%%%%%%%%% 
%%%%%%%%%%%%%%%%%%%%%%%%%%%%%%%%%%%%%%%%%%%%%%%%%% 
Next we consider a flat space ${\mathbb R}^n$ and a torus $T^n$.  
In this case, no tachyon appears, and a perturbative calculation is
possible without any constraint. 

The covariant derivative and the Lorentz generators are 
identified with the translation and rotation generators 
of the Poincar\'e  group $ISO(n)$. 
Therefore, the principal $Spin(n)$ bundle 
corresponding to ${\mathbb R}^{n}$ is the Poincar\'e  group:
\begin{eqnarray}
E_{prin}({\mathbb R}^n)=ISO(n). 
\end{eqnarray}
Compactifying $E_{prin}({\mathbb R}^n)$, we obtain $E_{prin}(T^n)$. 

Although $E_{prin}({\mathbb R}^n)$ has a group structure, 
for actual calculations it is better to use another regularization. 
The strategy is to use a momentum basis 
in the direction of ${\mathbb R}^n$ (or $T^n$), 
while in the direction of $Spin(n)$ we use a coordinate
basis\footnote{
  In such a basis, fields with various spins are mixed. 
} \cite{KK}. 
With such a basis, the covariant derivative can be written as 
\begin{eqnarray}
\left(\nabla_{(a)}\right)_{IJ}
=
ip_{(a)}^I\delta_{IJ}
=
i\left(R_{(a)}{}^b(g^{-1})\right)^{\alpha}
\left(p_b\right)^i\delta_{\alpha\beta}\delta_{ij}, 
\end{eqnarray}
where $i,j$ and $\alpha,\beta$ indicate the degrees of freedom 
of ${\mathbb R}^n$ and $Spin(n)$, respectively,  
and we have $I=(i,\alpha)$ and $J=(j,\beta)$. 
The quantity $p_b^i$ is an eigenvalue of $-i\partial_b$, 
which is the momentum in the direction of ${\mathbb R}^n$. 
We assume that $p_b^i$ is uniformly distributed below the momentum cutoff. 
Then, $p_{(a)}^I$ is also uniformly distributed. 
Therefore, by regarding $p_{(a)}$ as the 
``momentum along the $(a)$-th direction", 
we can apply the techniques of the quenched reduced model \cite{quench}. 
%%%%%%%%%%%%%%%%%%%%%%%%%%%%%%%%%%%%%%%%%%%%%%%%%%
%%%%%%%%%%%%%%%%%%%%%%%%%%%%%%%%%%%%%%%%%%%%%%%%%%
%%%%%%%%%%%%%%%%%%%%%%%%%%%%%%%%%%%%%%%%%%%%%%%%%%
\section{Conclusions}\label{sec:discussion}
%%%%%%%%%%%%%%%%%%%%%%%%%%%%%%%%%%%%%%%%%%%%%%%%%%
%%%%%%%%%%%%%%%%%%%%%%%%%%%%%%%%%%%%%%%%%%%%%%%%%%
%%%%%%%%%%%%%%%%%%%%%%%%%%%%%%%%%%%%%%%%%%%%%%%%%%
In this paper, we discussed a method for regularizing 
the covariant derivative on a Riemannian manifold 
by a set of finite-$N$ matrices. 
Such explicit constructions are important for various reasons. 
As an example, consider spontaneous spacetime generation \cite{AIKKT}. 
In our interpretation, the four-dimensional eigenvalue distribution 
observed in Ref. \citen{mf} suggests that the four-dimensional 
spacetime is realized dynamically. 
However, extensity of the eigenvalues does not reveal the topology
\footnote{This is in a sharp contrast to the original interpretation 
\cite{AIKKT}. }. 
(Note that any manifold entails infinitely large momentum.)
In order to extract information concerning the topology, 
we must study the detailed structure, 
such as  commutation relations and the degeneracy of eigenvalues. 

Although for generic manifolds it is difficult to write down 
explicit forms of regularizations, 
for some classes of manifolds 
that have higher symmetries, we can introduce explicit cutoffs, 
and explicit calculations are possible. 
We studied spheres, real and complex projective spaces, tori  
and flat spaces as examples. 

In the case of the $n$-dimensional sphere $S^n$ and 
the real projective space ${\mathbb R}P^n$, 
the corresponding principal $Spin(n)$ bundles are 
$Spin(n+1)$ and $SO(n+1)$, respectively, 
and the covariant derivative can be expressed in terms of 
the rotation generators acting on the regular representation.  
By discarding the larger representations, we can introduce a cutoff 
for both the momentum and spin of the fields. The meaning of the states in
the Hilbert space is clear. 
The covariant derivatives on $S^n$ and ${\mathbb R}P^n$ satisfy 
the same commutation relation, but they are expressed 
in terms of different
matrices; indeed, the degeneracies of the irreducible representations 
are different. This is a reflection of the fact that 
$S^n$ and ${\mathbb R}P^n$ are locally isomorphic 
but globally non-isomorphic. 

We investigated the case of $S^2$ in detail and 
found how the local fields and 
the generators of the diffeomorphism, the local Lorentz transformation, 
and the higher spin gauge transformations are embedded in the
components of the matrices. 
If we choose a Yang-Mills-type action, such as that of 
IIB matrix model, a perturbative expansion about this background 
cannot be carried out, due to the existence of tachyonic modes. 
Such modes come from the local fields, which depend on the coordinate
of $Spin(2)$. If we use an action that does not entail such tachyons, 
then using the techniques presented in \S \ref{sec:regularization}, 
the calculation becomes almost parallel to that on a fuzzy sphere background.  
 
In the case of a flat background, we can use the techniques of 
the quenched reduced models \cite{quench}. 
In this case, no tachyonic modes appear. 
Because the quenched reduced model reproduces the $U(N)$ gauge theory, 
by studying this background, we may find new connections between 
gauge theory and gravity \cite{HKK3}. 
A calculation of the correlators of the vertex operators 
\cite{VO} would be very useful in this context.  
It would also be useful to clarify the relation between our formalism and 
that for the original interpretation of IIB matrix model. 
%%%%%%%%%%%%%%%%%%%%%%%%%%%%%%%%%%%%%%%%%%%%%%%%%%
%%%%%%%%%%%%%%%%%%%%%%%%%%%%%%%%%%%%%%%%%%%%%%%%%%
%%%%%%%%%%%%%%%%%%%%%%%%%%%%%%%%%%%%%%%%%%%%%%%%%%
\section*{Acknowledgements}
%%%%%%%%%%%%%%%%%%%%%%%%%%%%%%%%%%%%%%%%%%%%%%%%%%
%%%%%%%%%%%%%%%%%%%%%%%%%%%%%%%%%%%%%%%%%%%%%%%%%%
%%%%%%%%%%%%%%%%%%%%%%%%%%%%%%%%%%%%%%%%%%%%%%%%%%
The author would like to thank H. Kawai and Y. Kimura  
for helpful discussions, comments and 
collaboration on related topics. 
He also thanks T. Hirata, F. Kubo and Y. Matsuo for discussions.  
This work was supported in part by JSPS Research Fellowships for Young Scientists. 
This work was also supported in part by a Grant-in-Aid for
the 21st Century COE ``Center for Diversity and Universality in
Physics''. 

\appendix
%%%%%%%%%%%%%%%%%%%%%%%%%%%%%%%%%%%%%%%%%%%%%%%%%%
%%%%%%%%%%%%%%%%%%%%%%%%%%%%%%%%%%%%%%%%%%%%%%%%%%
%%%%%%%%%%%%%%%%%%%%%%%%%%%%%%%%%%%%%%%%%%%%%%%%%%
\section{The Heat Kernel Regularization}\label{appendix:heat kernel}
%%%%%%%%%%%%%%%%%%%%%%%%%%%%%%%%%%%%%%%%%%%%%%%%%%
%%%%%%%%%%%%%%%%%%%%%%%%%%%%%%%%%%%%%%%%%%%%%%%%%%
%%%%%%%%%%%%%%%%%%%%%%%%%%%%%%%%%%%%%%%%%%%%%%%%%%
\hspace{0.51cm}
In this appendix, we present the formulae used in 
\S \ref{sec:heat kernel}. 
We start with the well-known formula 
\begin{eqnarray}
  Tr_t\ \textbf{1}
  &=&
  \kappa^{d-D}
  \int dg \int e d^dx
  \langle x,g| e^{t\Delta_{prin}}|x,g\rangle
  \nonumber\\
  &=&
  \frac{\kappa^{d-D}t^{-D/2}}{(4\pi)^{D/2}}
  \int dg \int e d^dx
  \left(
    1+\frac{t}{6}R_{prin}+O(t^2)
  \right), 
\end{eqnarray}
where $R_{prin}$ is the scalar curvature of $E_{prin}$, 
and $D=\frac{d(d+1)}{2}$ is the dimension of $E_{prin}$.  
At low energies, we can ignore all terms but the first 
in the limit $t\to 0$. 

For simplicity, let us denote both the subscript $(a)$ of $\nabla_{(a)}$ 
and $ab$ of ${\cal O}_{ab}$ by a single index, $I$, and denote 
$\nabla_{(a)}$ and $\kappa{\cal O}_{ab}$ by $D_I$.  
Then, we can evaluate the trace of $f^{IJ\cdots K}(x,g)D_I D_J\cdots
D_K$ (we assume $I,J,\cdots,K$ to be symmetric) as follows. 

By the general covariance, the trace of $f^{IJ}(x,g)D_I D_J$ takes the
form 
\begin{eqnarray}
  \lefteqn{
    Tr_t\left(
      f^{IJ}(x,g)D_I D_J
    \right)
  }\nonumber\\
  &=&
  \kappa^{d-D}t^{-D/2-1}
  \int dg \int e d^dx
  \left(
    \alpha \delta_{IJ}f^{IJ}
    +
    \beta t R_{prin}f^{IJ}\delta_{IJ}
    +
    \gamma t (R_{prin})_{IJ} f^{IJ}
    +
    \cdots
  \right), 
  \nonumber\\
\end{eqnarray}
where $R_{prin}$ and $(R_{prin})_{IJ}$ are the scalar curvature 
and the Ricci tensor of $E_{prin}(M)$, respectively. 
The coefficients can be determined by taking $f^{IJ}=\delta^{IJ}$. 
For example, from the relations 
\begin{eqnarray}
  Tr_t\left(
    \delta^{IJ}D_I D_J
  \right)
  &=&
  Tr_t\left(
    \Delta_{prin}
  \right) 
  \nonumber\\
  &=&
  \frac{d}{dt}\left( Tr_t\ \textbf{1}\right)
  \nonumber\\
  &=&
  -\frac{D}{2}
  \frac{\kappa^{d-D}t^{-D/2-1}}{(4\pi)^{D/2}}
  \left(\int dg \int e d^dx\right)
  +
  \cdots  
\end{eqnarray}
and 
\begin{eqnarray}
  Tr_t\left(
    \delta^{IJ}D_I D_J
  \right)
  &=&  
  \kappa^{d-D}t^{-D/2-1}
  \int dg \int e d^dx
  \alpha \delta_{IJ}\delta^{IJ}
  +
  \cdots
  \nonumber\\
  &=&
  \alpha D
  \cdot
  \kappa^{d-D}t^{-D/2-1}
  \left(\int dg \int e d^dx\right)
  +
  \cdots, 
\end{eqnarray}
we have 
\begin{eqnarray}
  \alpha=-\frac{1}{2(4\pi)^{D/2}}. 
\end{eqnarray}
Similarly, we have 
\begin{eqnarray}
  \lefteqn{
    Tr_t\left(
      f^{IJKL}(x,g)D_I D_J D_K D_L
    \right)  
  }\nonumber\\
  &=&
  \frac{\kappa^{d-D}t^{-D/2-2}}{4(4\pi)^{D/2}}
  \int dg\int e d^dx
  f^{IJKL}
  \left(
    \delta_{IJ}\delta_{KL}
    +
    \delta_{IK}\delta_{JL}
    +
    \delta_{IL}\delta_{JK}
  \right)
  +
  \cdots, 
  \\
  \lefteqn{
    Tr_t\left(
      f^{IJKL}(x,g)D_I D_J D_K D_L D_M D_N
    \right)  
  }\nonumber\\
  &=&
  -\frac{\kappa^{d-D}t^{-D/2-3}}{8(4\pi)^{D/2}}
  \int dg\int e d^dx
  f^{IJKL}
  \left(
    \delta_{IJ}\delta_{KL}\delta_{MN}
    +
    \mbox{14 permutations}
  \right)
  +
  \cdots. 
  \nonumber\\
\end{eqnarray}
Note that terms with an odd number of $D_I$ vanish. 
%%%%%%%%%%%%%%%%%%%%%%%%%%%%%%%%%%%%%%%%%%%%%%%%%%
%%%%%%%%%%%%%%%%%%%%%%%%%%%%%%%%%%%%%%%%%%%%%%%%%%
%%%%%%%%%%%%%%%%%%%%%%%%%%%%%%%%%%%%%%%%%%%%%%%%%%
\section{Deformed Actions and Their Classical Solutions}
\label{sec:deformed action}
%%%%%%%%%%%%%%%%%%%%%%%%%%%%%%%%%%%%%%%%%%%%%%%%%%
%%%%%%%%%%%%%%%%%%%%%%%%%%%%%%%%%%%%%%%%%%%%%%%%%%
%%%%%%%%%%%%%%%%%%%%%%%%%%%%%%%%%%%%%%%%%%%%%%%%%%
\hspace{0.51cm}
As we saw in the Introduction, with the ansatz 
\begin{eqnarray}
  A_a=i\nabla_{(a)}, 
  \qquad
  \psi_\alpha=0, 
  \label{ansatz1}
\end{eqnarray}
the equation of motion of IIB matrix model gives 
Ricci flat spacetimes as classical solutions. 
Note that Ricci-flat spacetimes 
with fewer than $10$ dimensions are also 
classical solutions; indeed, the only difference is 
that in this case we use the ansatz 
\begin{eqnarray}
  A_a=i\nabla_{(a)}\ (a=1,\cdots,n), 
  \qquad
  A_a=0\ (a=n+1,\cdots,10), 
  \qquad
  \psi_\alpha=0. 
  \label{ansatz2}
\end{eqnarray}

By deforming the action, many noncommutative spaces become 
classical solutions. 
In this appendix we determine the kinds of classical solutions 
that exist in our interpretation.  
%%%%%%%%%%%%%%%%%%%%%%%%%%%%%%%%%%%%%%%%%%%%%%%%%%
%%%%%%%%%%%%%%%%%%%%%%%%%%%%%%%%%%%%%%%%%%%%%%%%%%
%%%%%%%%%%%%%%%%%%%%%%%%%%%%%%%%%%%%%%%%%%%%%%%%%%
\subsection{Matrix model with a mass term}
%%%%%%%%%%%%%%%%%%%%%%%%%%%%%%%%%%%%%%%%%%%%%%%%%%
%%%%%%%%%%%%%%%%%%%%%%%%%%%%%%%%%%%%%%%%%%%%%%%%%%
%%%%%%%%%%%%%%%%%%%%%%%%%%%%%%%%%%%%%%%%%%%%%%%%%%
If we add a mass term 
\begin{eqnarray}
  S_{mass}=-\frac{m^2}{g^2}
  Tr\left(
    A_{a}A^{a}
  \right) 
\end{eqnarray}
to (\ref{action_IIB}), 
the equation of motion becomes 
\begin{eqnarray}
  \left[
    A_{b},
    \left[
      A_{a},A^{b}
    \right]
  \right]
  +
  2m^2A_{a}
  =
  0. 
  \label{EOM:IIB with mass term}
\end{eqnarray}
Here we set $\psi=0$ for simplicity. 
Substituting the ansatz (\ref{ansatz1}) or (\ref{ansatz2}) 
into (\ref{EOM:IIB with mass term}), we have 
\begin{eqnarray}
  \nabla^a R_{ab}{}^{cd}=0, 
  \qquad
  R_{ab}=2m^2\delta_{ab}. 
  \label{classical solution:IIB with mass term}
\end{eqnarray}
Using the Bianchi identity, the former follows from the latter.  
Note that Eq. (\ref{classical solution:IIB with mass term}) is 
the Einstein equation with a cosmological constant 
\begin{eqnarray}
  \Lambda=(n-2)m^2. 
\end{eqnarray}
%%%%%%%%%%%%%%%%%%%%%%%%%%%%%%%%%%%%%%%%%%%%%%%%%%
%%%%%%%%%%%%%%%%%%%%%%%%%%%%%%%%%%%%%%%%%%%%%%%%%%
%%%%%%%%%%%%%%%%%%%%%%%%%%%%%%%%%%%%%%%%%%%%%%%%%%
\subsection{Matrix model with a cubic term}
%%%%%%%%%%%%%%%%%%%%%%%%%%%%%%%%%%%%%%%%%%%%%%%%%%
%%%%%%%%%%%%%%%%%%%%%%%%%%%%%%%%%%%%%%%%%%%%%%%%%%
%%%%%%%%%%%%%%%%%%%%%%%%%%%%%%%%%%%%%%%%%%%%%%%%%%
Next, let us consider the matrix model with the cubic term 
\cite{CS} 
\begin{eqnarray}
  S_{CS}
  =
  \frac{i\alpha}{3g^2}\epsilon^{abc}
  Tr\left(
    [A_a,A_b]A_c
  \right), 
\end{eqnarray}
which has the fuzzy sphere as a classical solution. 
Here $\epsilon^{abc}$ is totally antisymmetric, 
$\epsilon^{123}=1$, and $\epsilon^{abc}=0$ if at least one of the indices is
not $1,2$ or $3$.  
 
The equation of motion is 
\begin{eqnarray}
  \left[
    A_{b},
    \left[
      A_{a},A^{b}
    \right]
  \right]
  -
  i\alpha\epsilon^{abc}[A_b,A_c]
  =
  0, 
  \label{EOM:IIB with CS term}
\end{eqnarray}
and using the ansatz (\ref{ansatz1}) or (\ref{ansatz2}), 
we obtain 
\begin{eqnarray}
  R_{ab}{}^{cd}=0
\end{eqnarray}
for $n=2,3$ and 
\begin{eqnarray}
  \nabla^b R_{ab}{}^{de}
  -
  \alpha\epsilon^{abc}
  R_{bc}{}^{de}
  =
  0, 
  \qquad
  R_{ab}=0
\end{eqnarray}
for $n\ge 4$. 
Using the Bianchi identity, we can rewrite the latter as 
\begin{eqnarray}
  R_{ab}=0, 
  \qquad
  \epsilon^{abc}
  R_{bc}{}^{de}
  =
  0. 
\end{eqnarray}
In this way, it is found that with the ansatz (\ref{ansatz1}) and 
(\ref{ansatz2}), classical solutions of the matrix model 
with the Chern-Simons term are rather trivial. 
%%%%%%%%%%%%%%%%%%%%%%%%%%%%%%%%%%%%%%%%%%%%%%%%%%%%%%%
%%%%%%%%%%%%%%%%%%%%%%%%%%%%%%%%%%%%%%%%%%%%%%%%%%%%%%%

\end{document}